# Global Dormancy of Metastases Due to Systemic Inhibition of Angiogenesis

Sébastien Benzekry[1,2], Alberto Gandolfi[3], Philip Hahnfeldt[1]*

1 Center of Cancer Systems Biology, GRI, Tufts University School of Medicine, Boston, Massachusetts, United States of America, 2 Inria team MC2, Institut de Mathématiques de Bordeaux, Bordeaux, France, 3 Istituto di Analisi dei Sistemi ed Informatica "Antonio Ruberti", Roma, Italy

**Abstract**

Autopsy studies of adults dying of non-cancer causes have shown that virtually all of us possess occult, cancerous lesions. This suggests that, for most individuals, cancer will become dormant and not progress, while only in some will it become symptomatic disease. Meanwhile, it was recently shown in animal models that a tumor can produce both stimulators and inhibitors of its own blood supply. To explain the autopsy findings in light of the preclinical research data, we propose a mathematical model of cancer development at the organism scale describing a growing population of metastases, which, together with the primary tumor, can exert a progressively greater level of systemic angiogenesis-inhibitory influence that eventually overcomes local angiogenesis stimulation to suppress the growth of all lesions. As a departure from modeling efforts to date, we look not just at signaling from and effects on the primary tumor, but integrate over this increasingly negative global signaling from all sources to track the development of total tumor burden. This *in silico* study of the dynamics of the tumor/metastasis system identifies ranges of parameter values where mutual angio-inhibitory interactions within a population of tumor lesions could yield global dormancy, i.e., an organism-level homeostatic steady state in total tumor burden. Given that mortality arises most often from metastatic disease rather than growth of the primary *per se*, this finding may have important therapeutic implications.

**Citation:** Benzekry S, Gandolfi A, Hahnfeldt P (2014) Global Dormancy of Metastases Due to Systemic Inhibition of Angiogenesis. PLoS ONE 9(1): e84249. doi:10.1371/journal.pone.0084249

**Editor:** Aamir Ahmad, Wayne State University School of Medicine, United States of America

**Received** August 20, 2013; **Accepted** November 13, 2013; **Published** January 21, 2014

**Copyright:** © 2014 Benzekry et al. This is an open-access article distributed under the terms of the Creative Commons Attribution License, which permits unrestricted use, distribution, and reproduction in any medium, provided the original author and source are credited.

**Funding:** This work was supported by the National Cancer Institute under Award Number U54CA149233 (to L. Hlatky). The funders had no role in study design, data collection and analysis, decision to publish, or preparation of the manuscript.

**Competing Interests:** The authors have declared that no competing interests exist.

* E-mail: philip.hahnfeldt@tufts.edu

## Introduction

Almost all of us carry small tumor lesions that for many will not progress to symptomatic disease. Indeed, as evidenced in autopsy studies for adults without pre-established cancer such as [1,2], occult lesions are present in most healthy adults. Nielsen et al. [3] found that, out of 110 women cases, among which only one had been previously treated for breast cancer, 22% had at least one malignant lesion. Moreover, 45% of these had multicentric lesions. Similar results have been reported for prostate cancer in men [4]. For thyroid cancer, autopsy results [2] showed a prevalence rate of 99.9% for occult carcinomas, while incidence of thyroid cancer is only 0.1% [5].

To explain these results, it is necessary to understand the tumor dormancy phenomenon. Tumor dormancy [6,7] is defined by stable or very slow tumor growth. It can happen at the cellular level as a malignant cell remaining quiescent for a long period before awakening, but here we focus on the mm-scale lesions such as have surfaced in the several remarkable autopsy studies discussed, i.e., tissue-level tumor dormancy. Although the sizes of these dormant tumors remain almost constant, it is not due to a cessation in cell proliferation, but rather to increased apoptosis that leads to a near zero net growth rate [6–8]. Clinically, tumor dormancy has been observed in breast cancer [3,9–11], melanoma [12] and prostate cancer [4], among many others [6]. Dormancy is particularly relevant to the situation where secondary tumors (metastases) remain small and undetectable for extended periods.

Various explanations have been proposed for tumor dormancy, among these being the achievement of a balance between stimulation and inhibition of angiogenesis [7,13,14]. This mechanism offers one explanation for how secondary tumors may be suppressed to a near-dormant state by the primary; a phenomenon known as 'concomitant resistance' [15,16]. In fact, a number of explanations for the concomitant resistance phenomenon have been suggested, as well summarized by Chiarella et al. [16]: 1) monopolization of certain resources by the primary tumor that deprives secondary tumors of materials needed for growth, 2) primary tumor-induced enhancement of immune suppression of small secondary tumors (concomitant immunity), 3) anti-proliferative molecules released by the primary tumors and 4) release of angiogenesis inhibitors by the primary tumor into the blood circulation resulting in inhibition of vascular development at secondary sites. Nevertheless, although a distant impairment of metastatic growth by a primary tumor has been recognized for over a hundred years [17], and has meanwhile been informed by various preclinical [18–22] and clinical [9,23–25] studies, it remains poorly understood.

However, because of evidence that concomitant resistance happens in immune-deficient mice [21] and considering the large and unequivocal body of support for the role angiogenesis inhibition plays in the maintenance of tumor dormancy [8,26–30] and the "angiogenic switch" [31] in escape from dormancy, our focus here will be on the last theory. Angiogenesis, the process of creating new blood vessels and developing a supporting vascular network, was shown by Folkman [32] to be critical for tumor growth. Indeed, without development of new blood vessels, a malignant neoplasm cannot grow further than about 2 to 3 mm in diameter, due to nutrient supply limitations [32]. This process is regulated by the release from cancer cells of stimulatory growth factors, such as vascular endothelial growth factor (VEGF), that induce proliferation, migration and maturation of surrounding endothelial cells, as well as the production of angiogenesis inhibitory factors that act to curtail endothelial expansion [33]. As an example, in 1994 when examining the growth of Lewis lung carcinoma in a syngeneic murine tumor model, O'Reilly et al. [26] discovered an endogenous molecule having an inhibitory effect on angiogenesis, which they called 'angiostatin', followed soon by the discovery of 'endostatin' [27]. Endogenous anti-angiogenic molecules were also evidenced in human cancer, an example being thrombospondin-1 [29]. Overlaying the ability of tumors to stimulate vasculature, the discovery of their ability to also inhibit it [34] allows for the possibility that tumors may indirectly control their own growth [14,33,34], perhaps as a vestige of normal organ growth control. Further, inherent to this self-control notion, if the inhibitors were longer-lived and thus more persistent in the circulation, they could have the collateral effect of suppressing angiogenesis and growth at distant metastatic sites as the tumor mass gets large [14]. Indeed, the half-life of angiogenesis stimulators has been reported to be on the order of minutes for VEGF [35], while that for angiogenesis inhibitors is on the order of hours [26,29].

Amidst these developments, there have been a number of efforts to take numerous complex mechanisms of cancer biology into account in mathematical models (see [36] for a review), but very few of these models have had the aim of describing metastatic development, despite metastasis being the main cause of death from cancer [37]. Indeed, while the cure rate of cancer before appearance of metastases is about 90% for all cancers combined, it falls to just 15% when distant metastases are present at diagnosis [16]. As far as we know, modeling efforts in this direction can only be found in the work of Liotta and coworkers [38], and more recently in a few stochastic models [39–42] describing progression through the different stages of the metastatic process (cell detachment, intravasation, survival in the blood, extravasation, settling in a new environment), and in one notable dynamic model [43]. In this last case, Iwata and coworkers [43] proposed a quantitative formalism for the development of metastatic colonies, which was of great potential interest as it is was designed to describe the size distribution of the metastases, allowing thus to distinguish between micro-metastases and larger lesions. However, this model does not take angiogenesis into account. We therefore decided to theoretically combine this work with that of Hahnfeldt et al. [14] for tumor development under angiogenic control, along with a mathematical model developed for the growth and dissemination of a metastatic population [44,45]. The goal we realized was a new global formalism that integrates local stimulation with systemic inhibition of angiogenesis by a circulating factor produced by each lesion in a population of tumors, to provide insight into the development of the entire tumor/metastasis system.

## Methods and Results

### In silico model – derivation and implementation

The global philosophy of the model we propose is to consider the development of cancer disease at the organism scale, by describing the colonization and dissemination of a population of secondary tumors (metastases), in parallel with the growth of the primary lesion, taking into account organism-scale signaling interactions amongst these various tumor sites. The impetus for this viewpoint comes from Iwata et al. [43] where the authors derived a structured population model for describing the metastatic colonies represented by a density structured in size (volume). This model consists of a linear transport partial differential equation with a nonlocal boundary condition of renewal type. It has been further mathematically studied in [46,47], in particular to develop efficient numerical methods for discretizing the problem.

A major limitation of this approach, though, is that it does not take angiogenesis into account, although this is a fundamental process of tumor development that cannot be neglected, particularly if we want to study the effects of clinical angiogenesis inhibition. However, by combining the approach of Iwata et al. [43], arguably the first dynamical model for metastatic development, along with the model of Hahnfeldt et al. [14], which is the first to consider angiogenic homeostatic control of tumor growth, we developed in previous work a hybrid construct that integrates the angiogenic process into the growth of each tumor [44,45,48]. Since this model was written at the level of the organism, it was considered a suitable framework to adapt to the problem of analytically describing the consequences of systemic inhibition of angiogenesis (SIA).

The result is a model for tumor growth control that takes into account the local and systemic actions of angiogenesis regulators. It integrates the ability of tumor lesions to locally stimulate angiogenesis while simultaneously inhibiting angiogenesis globally, and is fitted to preclinical data. Information on the behavior of metastases is inferred from the estimated parameters. Simulations of the cancer history are performed, which provide a detailed description of the distribution of predicted metastatic lesion sizes. The biological hypothesis of a global dormancy state of self-inhibiting tumors is then tested, and corresponding ranges of the inhibitor production rate identified.

A schematic view of the new formalism we propose is presented in Figure 1. The main feature added to the previous model (Benzekry [48]) is a new variable representing the circulating concentration of an endogenous angiogenesis inhibitor, standing in for all possible inhibitory molecules (examples being endostatin, angiostatin or thrombospondin-1) impacting on the growth of each tumor. As a general modeling principle, we sought to be parsimonious and describe the major dynamics of the system with as few parameters as possible to assure each dynamic introduced carries its proper burden to explain the data.

### Mathematics of tumor growth and systemic inhibition of angiogenesis

Our construct considers primary tumors and their metastases to be distinct lesions whose states are described by two traits: volume $V$ and carrying capacity $K$. The primary tumor state is denoted $(V_p(t), K_p(t))$. The model's main variable is $\rho(t, V, K)$, the physiologically structured density of metastases having volume $V$ and carrying capacity $K$ at time $t$. The term density means that the metastases are assumed to exist in a continuum of sizes and carrying capacities and that the number of tumors between volumes $V_1$ and $V_2$ and carrying capacities between $K_1$ and $K_2$ is given by $\int_{V_1}^{V_2} \int_{K_1}^{K_2} \rho(t,V,K) dV dK$. We assume that the dynamics of each tumor's state are governed by a growth rate for $(V, K)$, denoted by the vector $G(V, K; V_p, \rho)$, that dissemination of new metastases is driven by a volume-dependent emission rate $\beta(V)$, and that the repartition of metastases at birth is given by $\mathcal{N}(V, K)$.

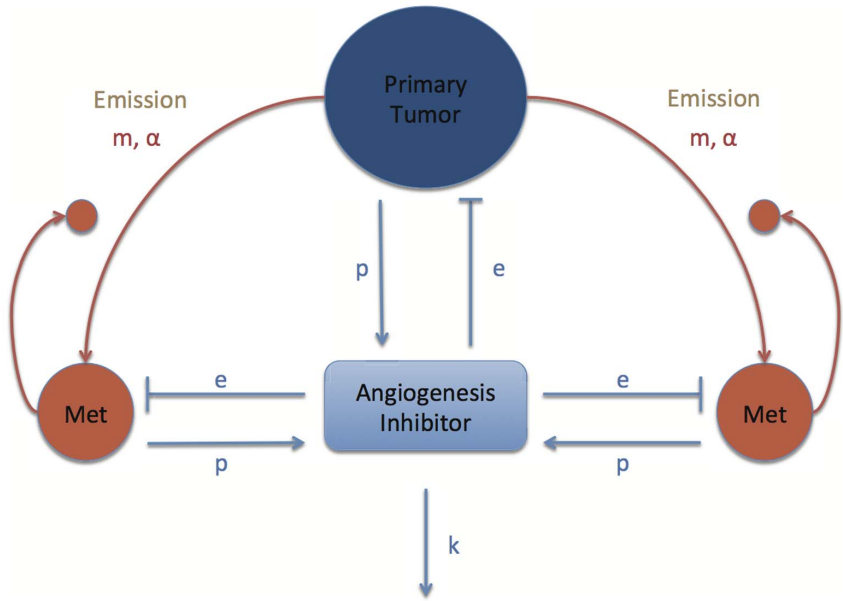

**Figure 1. Schematic representation of the model for systemic inhibition of angiogenesis.** *m*, *α*: metastatic spreading parameters. *p*: production rate of angiogenesis inhibitor. *e*: efficacy parameter of inhibitor. *k*: elimination rate of the inhibitor.
doi:10.1371/journal.pone.0084249.g001

The precise expressions of these functions will be described below. We consider some fixed final time $T$ and a physiological domain $\Omega$ for the possible values of $(V, K)$, defined as $\Omega = (V_0, +\infty) \times (0, +\infty)$ where the distribution of metastases has its support, which means that metastases have size bigger than the size of one cell $V_0$ and non-negative carrying capacity. In the formula below, the vector $v(V, K)$ stands for the external unit normal to the boundary $\partial\Omega$ of the domain $\Omega$. The notation $\partial\Omega^+$ stands for the subset of the boundary where the flux is pointing inward, i.e. where $G(V, K; V_p, \rho) \cdot v(V, K) < 0$. The map $(V, K) \mapsto \rho^0(V, K)$ denotes the initial distribution of the metastatic colonies.

Overall, the model we arrived at is a nonlinear transport partial differential equation of renewal type with a nonlocal boundary condition.

$$\left.\begin{array}{l}\partial_t \rho + \text{div}(\rho G) = 0 \quad\quad\quad\quad\quad\quad (0,T) \times \Omega \\ -G(V,K;V_p,p) \bullet v(V,K)\rho(t,V,K) = N(V,K) \\ \left\{\int_\Omega \beta(V)\rho(t,V,K)dVdK + \beta(V_p(t))\right\} \quad (0,T) \times \partial\Omega^+ \\ \rho(0,V,K) = \rho^0(V,K) \quad\quad\quad\quad\quad\quad \Omega\end{array}\right\} \quad (1)$$

We now make precise the expressions of the various coefficients of the model; in particular how the growth rate $G$ is affected by the total population of tumors represented by $\rho$. We assume that all the tumors (primary and secondaries) share the same growth model but have different parameters, due to the different sites where they are located. However, within the population of metastases, all tumors are assumed to grow with the same parameters. The growth velocity of each tumor is given by a vector field $G(V, K; V_p, \rho)$. Following the approach of [14] we assume

$$G(V,K;V_p,\rho) = \begin{pmatrix} aV\ln\left(\frac{K}{V}\right) \\ Stim(V,K) - Inhib(V,K;V_p,\rho) \end{pmatrix}.$$

In the previous expression, the first line is the rate of change of the tumor volume $V$ (where $a$ is a constant parameter driving the proliferation kinetics of the cancer tissue) and the second line is the rate of change of the carrying capacity $K$. The main idea of this tumor growth model is to start from a gompertzian growth of the tumor volume (or any carrying capacity-like growth model [49]) and to assume that the carrying capacity $K$ is a dynamical variable representing the tumor environment limitations (here limited to the vascular support) changing over time. The balance between a stimulation term $Stim(V, K)$ and an inhibition term $Inhib(V, K; V_p(t), \rho(t, V, K))$ governs the dynamics of the carrying capacity. For the stimulation term we follow [14] and assume

$$Stim(V, K) = bV,$$

where the parameter $b$ is related to the concentration of angiogenic stimulating factors such as VEGF. This last quantity was derived to be constant in [14] from the consideration of very fast clearance of angiogenic stimulators [35].

For the inhibition term, Hahnfeldt et al. [14] only considered a local inhibition coming from the tumor itself. Our main modeling novelty is to consider in addition a global inhibition coming from the release in the circulation of angiogenic inhibitors by the total (primary + secondary) population of tumors. The following is an extension of the biophysical analysis performed in [14]. Let us consider a spherical tumor of radius $R$ inside the host body. The host is represented, for simplicity, by a single compartment of volume $V_d$ in which concentrations are assumed spatially uniform. Let $n(r)$ be the inhibitor concentration inside the tumor at radial distance $r$. Let the intra-tumor clearance of inhibitors, known to be slow, be (approximately) zero [14]. At quasi steady state, $n(r)$ then solves the following diffusion equation:

$$n''(r) + \frac{2n'(r)}{r} + \frac{p}{D^2} = 0,$$

where $p$ is the inhibitor production rate and $D^2$ is the inhibitor diffusion constant. This equation has the boundary condition $n(R) = i(t; V_p, \rho(t, V, K))$, where the expression on the right represents the systemic concentration of the inhibitor resulting from a primary tumor volume $V_p$ and secondary tumors of density $\rho$ at time $t$. Solving this equation (using that $n(0) < +\infty$) we obtain

$$n(r) = i + \frac{p}{6D^2}(R^2 - r^2).$$

From this expression we compute the mean inhibitor concentration in the tumor to obtain

$$Inhib(V, K; V_p, p) = \hat{e}\left(i + \frac{p}{15D^2}R^2\right)K =$$
$$\hat{e}\left(i + \frac{p}{15D^2}\left(\frac{3}{4\pi}\right)^{\frac{2}{3}}V^{\frac{2}{3}}\right)K,$$

where $\hat{e}$ is a sensitivity coefficient. For $i(t; V_p, \rho(t, V, K))$, considering that the total flux of inhibitors produced by a tumor with volume $V$ is $pV$ and assuming that the inhibitor production rate is the same in all the tumors, we have

$$V_d \frac{di}{dt} = pV_p + \int_\Omega pV\rho(t, V, K)dVdK - kV_d i,$$

where $k$ is an elimination constant from the blood circulation. Setting $I(t; V_p, \rho(t, V, K)) = V_d i(t; V_p, \rho(t, V, K))$, we get

$$\frac{dI}{dt} = pV_p + \int_\Omega pV\rho(t, V, K)dVdK - kI,$$

which has an initial condition that in significant cases may be set to zero, i.e., $I(t=0) = 0$. Overall, the explicit expression of the metastases growth rate is given by

$$G(V, K; V_p, \rho) = \begin{pmatrix} aV\ln\left(\frac{K}{V}\right) \\ bV - dV^{\frac{2}{3}}K - eIK \end{pmatrix}, \quad (2)$$

where $e = \dfrac{\hat{e}}{V_d}$ and

$$d = eV_d \frac{p}{15D^2}\left(\frac{3}{4\pi}\right)^{\frac{2}{3}}. \quad (3)$$

Note that we retrieve here the local term $dV^{2/3}$ from the analysis of [14]. Our analysis results in an additional global term $eI$ that captures the effect of systemic inhibition of angiogenesis.

For the primary tumor, we assume the same structural growth model. The dynamics of $(V_p, K_p)$ are thus given by

$$\frac{d}{dt}\begin{pmatrix} V_p \\ K_p \end{pmatrix} = G_p(V_p, K_p; V_p, \rho(t, V, K)) \quad (4)$$

where $G_p$ has the same expression as $G$, except that the parameters $a_p$ and $b_p$ (the values of $a$ and $b$ that are associated with the function $G$ for the primary) may be different from $a$ and $b$ associated with metastases, in those cases where the primary and metastases are presumed to have different growth kinetics. The inhibitor production rate $p$ and effect of the inhibitor $e$ are assumed to be the same for the primary and secondary tumors, which implies same value also for $d$ in view of formula (3).

### Metastatic dissemination

There is no clear consensus in the literature about metastases being able to metastasize or not [50–52]. However, we argue here that cancer cells that have acquired the ability to metastasize should conserve it when establishing a new site. Moreover, since metastases remain undetectable for an extended time [50–52] (in particular because tumors could remain dormant for some time), the absence of clear proof in favor of metastases from metastases could be due to the short duration of the experiments compared to the time required for a second generation of tumors to reach a visible size. Here we are interested in long-time behaviors and, although metastases from metastases could be neglected to a first approximation, we think this second-order term is relevant in our setting and chose to include it in our modeling, in light of some clinical evidence supporting second-generation metastases [53].

Successful metastatic seeding results when one malignant cell is able to overcome various adverse events including: detachment from the tumor, intravasation, survival in the blood/lymphatic circulation, escape from immune surveillance, extravasation, survival in a new environment (see [54] for more details about the biology of the metastatic cascade). Here, we regroup all these events into one emission rate $\beta(V, K)$, quantifying the number of successfully newly created metastases per unit of time. We assume very small metastases do not metastasize because they do not have access to the blood circulation, accounted for here by including a threshold $V_m$ below which tumors do not spread new individuals. $V_m$ is taken here to be 1 mm$^3$ as an approximation of the volume at which the angiogenic switch happens [32]. Apart from the addition of this threshold, the expression chosen for $\beta$ is the same as that used by Iwata et al. [43]:

$$\beta(V, K) = \beta(V) = \begin{cases} mV^\alpha & \text{if } V \geq V_m \\ 0 & \text{otherwise} \end{cases} \quad (5)$$

where $m$ and $\alpha$ are coefficients quantifying the overall metastatic aggressiveness of the cancer disease. The parameter $m$ represents the intrinsic metastatic potential of the cancer cells, and $\alpha$ represents the microenvironmental component of metastatic dissemination. It lies between 0 and 1 and is the third of the fractal dimension of the tumor vasculature, assumed here to be the same for all tumors. For instance, if vasculature develops superficially, then $\alpha = 2/3$, whereas for a fully penetrating vasculature, the value would be $\alpha = 1$. We here assume the dissemination rate depends only on the volume because simulations revealed that adding a monotone dependence on $K$ did not improve the flexibility of the model even while adding at least one parameter, contrary to the parsimony principle.

Stating a balance law for the number of metastases when they are growing in size gives the first equation of (1). The boundary condition, i.e. the second equation of (1), states that the entering flux of tumors equals the newly disseminated ones. These result from two sources: spreading from the primary tumor, modeled by the term $\beta V_p(t)$, and second-generation tumors coming from the metastases themselves, described by the term $\int_\Omega \beta(V)\rho(t,V,K)dVdK$. The map $(V,K) \mapsto N(V,K)$, where $(V,K) \in \partial\Omega$, stands for the volume- and carrying-capacity-dependent distribution of metastases at birth. Assuming that newly created tumors all have the size of 1 cell, denoted by $V_0$, and some initial carrying capacity, denoted by $K_0$, we have

$$N(V, K) = \delta_{(V,K)=(V_0, K_0)} \quad (6)$$

i.e. the Dirac distribution centered in $(V_0, K_0)$. We have previously discussed how this form can be deduced by passing to the limit from an absolutely continuous density [55]. An important feature of this model, in contrast to previous no-SIA models, is that we allow metastases to exit the domain by imposing the boundary condition only where the flux points inward and letting tumors exit the domain in the opposite situation. In view of expression (2), this occurs when the carrying capacity $K$ is less than the volume of one cell, $V_0$, i.e. when global inhibition is strong enough so that tumors can cross the line $K = V_0$, which is the case when $G(V_0, V_0) \cdot (0,1) = bV_0 - dV_0^{\frac{2}{3}} - eIV_0 < 0$. These tumors are then removed from the population, corresponding to death caused by nutrient deprivation.

From the solution $\rho$ of the model (1,3–6), biologically relevant macroscopic quantities can be defined, such as the total number of metastases $N(t) = \int_\Omega \rho(t,V,K)dVdK$, the total metastatic burden $M(t) = \int_\Omega V\rho(t,V,K)dVdK$, or the mean size of the metastases $\frac{M(t)}{N(t)}$.

### Solution-finding

To approximate the solutions of the problem (1,3–6) we adapted a numerical procedure previously developed for the model without SIA in [45,55]. It is a Lagrangian scheme based on the straightening of the characteristics of the transport equation. We then used an Euler method for discretization of the characteristics and computation of the primary tumor ordinary differential equation. The integral in the boundary condition was computed using the trapezoid approximation method.

### Parameters surmised from existing preclinical data

Data on metastatic development are not common in the literature, especially for micrometastases or dormant tumors, since these measurements are technically difficult to obtain. Even more difficult to find are data quantifying systemic inhibition of angiogenesis. For our purpose we use data from Huang et al. [56] that do not explicitly deal with systemic inhibition of angiogenesis nor global dormancy, but where number and mean size of metastases at the end time (T = 32 days) are available, together with primary tumor growth kinetics. The cell line used in this work is a spontaneous mouse breast cancer line 4T1, known to be highly metastatic with relatively slow primary tumor growth. Cells were injected subcutaneously ($10^5$ cells) in BALB/c mice. As shown in the following, our model was able to explain these experimental data.

Values of the parameters were fixed either by direct extraction from the literature, heuristic derivation, or by fitting the model to the data from [56]. For the preclinical data that we used, metastases actually develop to symptomatic volumes and did not evidence manifest global inhibition. Hence for fitting of the model, we consider SIA as being negligible and take $I = 0$. The parameter estimation we performed here is only intended for estimation of growth and metastatic spreading parameters. The assumption of negligible SIA was found *a posteriori* to be adequate for description of the data from [56], because adding an SIA term with reasonable parameter values did not have any impact on the model simulations, in the framework of the experiment from [56]. In the context of no SIA, there is no impact of the metastases on the primary tumor and we could separately fit the primary tumor growth and the metastatic development. This approach (compared to a global fitting of all the parameters together) further reduces the parameterization of the model and allows for more stable and biologically relevant parameter estimation. Indeed, only two degrees of freedom were used to fit the primary tumor growth (seven time points) and two for the data on metastases (two measurements). The meaning, units and values of the model parameters resulting from the whole estimation procedure are summarized in Table 1.

In Hahnfeldt et al. [14], values for the elimination rates $k$ and efficacy constants $e$ for two endogenous angiogenesis inhibitors, endostatin and angiostatin, were estimated by fitting tumor growth data of mice that received injections of these anti-angiogenic agents. We focus here on angiostatin, and use values for the agent efficacy $e$ and the elimination rate $k$ from the blood circulation reported in Hahnfeldt et al. [14], applying these to a 20 g mouse. This value gives a half life for angiostatin of 1.8 days, which is consistent with the value of 2.5 days that can be found in the literature [26].

O'Reilly et al. [26] showed that injection of 12.5 μg per day of recombinant human angiostatin reproduces the systemic inhibition due to a primary tumor removed when it reached the size of 1500 mm$^3$. An approximation of the production rate in their setting is $p \approx \frac{12.5}{1500} \times 10^{-3} \approx 8.3 \times 10^{-6}$ mg·mm$^{-3}$·day$^{-1}$. For the value of $V_d$ we argue that the blood volume of a mouse is about 1.2–1.6 cm$^3$ per 20 g body weight. Taking an approximate value of 1.4 cm$^3$ and assuming that the interstitial (extracellular) space fills 30% of the extravascular space (in agreement with measurements of the fraction of volume occupied by cells), summing the interstitial space and blood volume gives 6.98 cm$^3$. Hence we took $V_d = 7000$ mm$^3$ to be an approximation of the distribution volume. For the diffusion coefficient of angiostatin, $D^2$, we used a value 1.56 mm$^2$day$^{-1}$ taken from the literature [57]. Based on these values and the formula (3) for $d$ derived in the modeling section, we were able to heuristically compute an approximation of the parameter $d$ as $d \approx 0.0717$ mm$^{-2}$day$^{-1}$. In the following, we fixed $d$ and $d_p$ to this value, which allowed us to reduce possible indeterminacy in the parameter estimation for the growth model.

When reproducing the experiment of Huang et al. [56], we fixed the initial size of the primary tumor to be $V_p(0) = 0.1$ mm$^3$ (corresponding to $10^5$ cells, i.e. the number of cells injected into the mouse) and arbitrarily set the initial carrying capacity of the primary tumor to $K_p(0) = 200$ mm$^3$. Metastases were assumed to start with initial size $V_0 = 1$ cell $= 10^{-6}$ mm$^3$ and initial carrying capacity $K_0 = 1$ mm$^3$ (assumed to be an approximation of the maximum reachable size without angiogenesis [32]). For metastatic emission, we considered a superficial vascular development and took $\alpha = 2/3$, following what was estimated in Iwata et al. [43] from clinical data.

### Fits to the data

Parameters $a_p$ and $b_p$ were obtained by minimizing the sum of squared errors between the tumor growth model simulation and the primary tumor growth data from [56]. Least squares

**Table 1.** Values, units and meaning of the model parameters.

| Parameter | Value | Unit | Meaning | Rationale |
|---|---|---|---|---|
| $a_p$ | 0.154 | day$^{-1}$ | PT cells proliferation | Fit PT |
| $b_p$ | 16.7 | day$^{-1}$ | PT angiogenic stimulation | Fit PT |
| $d_p$ | 0.0717 | mm$^{-2}$ day$^{-1}$ | PT angiogenic local inhibition | H |
| $a$ | 0.154 | day$^{-1}$ | Met cells proliferation | Fit Met |
| $b$ | 12.5 | day$^{-1}$ | Met angiogenic stimulation | Fit Met |
| $d$ | 0.0717 | mm$^{-2}$ day$^{-1}$ | Met angiogenic local inhibition | H |
| $m$ | 0.0229 | mm$^{-3}$day$^{-1}$ | Colonization rate | Fit Met |
| $\alpha$ | 2/3 | | Fractal dimension of vascularization | [43] |
| $p$ | $8.3 \times 10^{-6}$ | mg mm$^{-3}$ day$^{-1}$ | Production of I | [26] |
| $k$ | 0.38 | day$^{-1}$ | Elimination rate of I | [14] |
| $e$ | 7.5 | mg$^{-1}$day$^{-1}$ | Effect of I | [14] |
| $V_0$ | $10^{-6}$ | mm$^3$ | Met initial volume | H |
| $K_0$ | 1 | mm$^3$ | Met initial carrying capacity | [32] |
| $V_m$ | 1 | mm$^3$ | Threshold for metastatic emission | H |
| $D^2$ | 0.156 | mm$^2$day$^{-1}$ | Angiostatin diffusion coefficient | [57] |
| $V_d$ | 7000 | mm$^3$ | Distribution volume | H |

PT = Primary Tumor. Met = metastases. I = global amount of angiogenic inhibitor in the blood. H = heuristic derivation.
doi:10.1371/journal.pone.0084249.t001

minimization was performed using the trust region reflective algorithm implemented in Matlab (Matlab 2009b, The Mathworks Inc.). We obtained good agreement between the fit and the data (Figure 2). Goodness of fit quantification by the $R^2$ value ($R^2 = 1 - \frac{\sum (y_i - f(t_i))^2}{\sum (y_i - \bar{y})^2}$, where the $y_i$ are the data points, $\bar{y}$ is the mean value of the data and the $f(t_i)$'s are the values of the model at times $t_i$) gave an excellent score of $R^2 = 0.99$.

Assuming that differences in growth between the primary tumor and its metastases should arise from interactions with the microenvironment, we fixed the proliferation parameter $\alpha$ for the metastases to the value obtained for the primary tumor growth. The only remaining parameters to be fixed were then $b$ (driving the angiogenic stimulation) and $m$ (controlling metastatic dissemination), allowing us to minimize the overall parameterization of the model (two parameters for two data points). These last two coefficients were determined by fitting the model to the experimental metastatic data of Huang et al. [56], the results of which are reported in Table 2. We obtained good agreement to the number and mean size of metastases. It was determined that with the estimated value of $m$ a tumor of 200 mm$^3$ spreads a new metastasis every 0.77 days.

The parameter estimation that we performed allowed us to simulate the experiment of [56] by using the parameters resulting from the model's fit (and $I = 0$). This gave further insight beyond the mere availability data could provide on the time development dynamics of the metastases and their final size distribution. Figure 3.A shows the growth in time of the total metastatic burden, while Figure 3.B depicts the colony size distribution at T = 32 days for an *in silico* replicate of the experiment performed in [56]. It reveals a nontrivial size distribution of the final metastatic colonies with a mode between 0.01 and 0.1 mm$^3$, and only one tumor with size larger than 10 mm$^3$. At this time the total lung metastatic burden is 63.5 mm$^3$ distributed between 48.5 tumors. Simulation performed with a non-zero $I$ and a value for $p$ extracted from [26] (see above) presented no significant difference in this setting compared to the simulation with $I = 0$, hence justifying *a posteriori* our assumption of negligible effect of SIA in the setting of [56].

These simulations and parameter estimation show that our mathematical model is a possible theory describing growth and development of primary and secondary tumors in a 4T1 cell line model. While in this context, systemic inhibition of angiogenesis had no significant importance in the small time range, it was concluded more broadly that the model, endowed with adequate parameter values, should be a useful theoretical tool for investigating a range of situations beyond the experimental setting of [56].

### Simulation of the cancer history from the first cancer cell predicts uncontrolled metastatic burden

Using our model and based on the parameters estimated in the previous section (Table 1), we were able to extrapolate to a totally

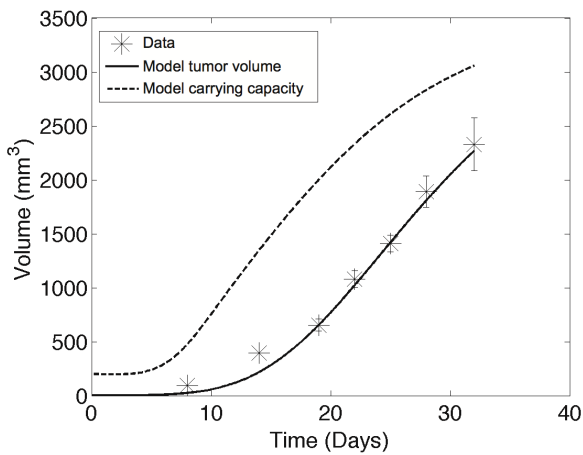

**Figure 2. Primary tumor growth.** Comparison of the fit of the model and the data from [56]. Data are mean $\pm$ standard error.
doi:10.1371/journal.pone.0084249.g002

**Table 2.** Metastatic outputs.

|  | Value from [56] | Computed by the model |
|---|---|---|
| **Median number of metastases (range)** | 43 (4–107) | 43.03 |
| **Mean size of metastases in mm³ (range)** | 1.47 (1.30–1.66) | 1.476 |

Comparison of the fit of the model and the data from [56]. For the number of metastases, the reported model value is the number of tumors above a minimal visible size that we took to be 10 cells (tumors were counted using a dissecting microscope in [56]). Mean size was given as diameter in [56] and was converted here into volume using $V = \frac{\pi}{6} \times D \times w^2$, $w = \frac{3}{4} D$.
doi:10.1371/journal.pone.0084249.t002

new setting where the primary tumor starts with one cell instead of an already large number of cells (approximately $10^5$). In so doing, we were able to simulate the whole cancer history, starting from one initial cancerous cell (and initial carrying capacity of 1 mm³) until the metastatic burden reached 5000 mm³, a burden we considered potentially lethal for a mouse. The simulation predicted this would happen 62.7 days after the first primary tumor cancer cell. Time development of the primary tumor volume, metastatic burden, total number and mean size of metastases as well as inhibitor amount in the host are plotted in Figure 4.

Interestingly, the model simulation predicts that the metastatic burden would overcome the primary tumor mass, implying that the mouse would probably die from growth of its secondary lesions rather than from the initial tumor. This is consistent with the metastatically aggressive phenotype of the 4T1 cell line. Quantification of the number of metastases reveals a final number of about 217 tumors, lots of them being small (Figure 4C) and probably undetectable in an experimental setting. Simulation with the same set of parameters but neglecting the effect of SIA ($I = 0$) showed no detectable difference on this time frame. Significant changes are observed later on, for volumes that are not considered to be physiologically relevant. This confirms that for the 4T1 cell line, metastases do develop and do not exhibit global dormancy, even when SIA is present with the inhibitor production parameter value extracted from [26]. Thus, based on biologically relevant parameters, our simulation results suggest large growth of the metastatic burden for the 4T1 cell line when starting from the first cancer cell, with a final metastatic volume larger than the primary tumor.

### Higher production of systemic angiogenesis inhibitor could result in long-term stable global dormancy in a population of self-inhibiting metastases

The previous simulations used parameter values derived from experimental data of a situation were metastases do develop and grow, because this is the only case where metastases are measurable and data are available. However we are interested here in global dormancy and situations where the metastatic population could remain ultimately small. We postulate that this could happen when production of the angiogenesis inhibitor, represented by parameter $p$ in our model, is significantly higher. Simulation results plotted in Figure 5 were obtained using a value $p = 2.5 \times 10^{-4}$ mg·mm$^{-3}$day$^{-1}$, i.e., a value about 30 times that extracted from [26]. From our previous modeling analysis and formula (3), higher production of inhibitor also proportionally increases the local inhibition parameters $d$ and $d_p$. In the simulation reported in Figure 5, we kept all the other parameter values unchanged from Table 1 and fixed the initial primary tumor volume to $V_p(0) = 1$ cell and the initial primary tumor carrying capacity to $K_p(0) = 1$ mm³. We simulated the system over a time of 350 days, covering the estimated lifespan of a mouse after appearance of an initial malignant cell. We focused on asymptotic behavior and possible convergence of the system to a steady state.

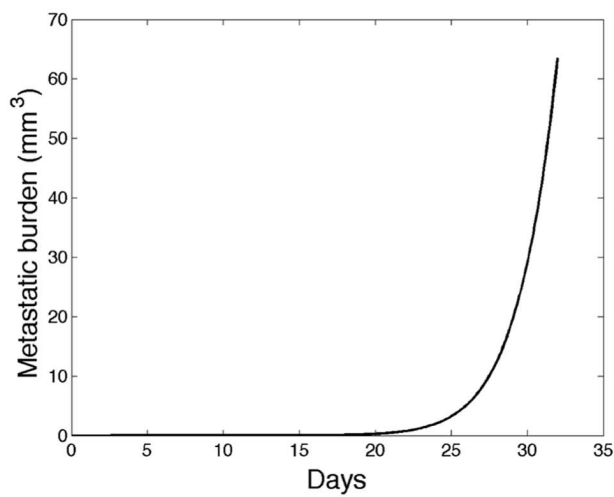

A. Metastatic burden

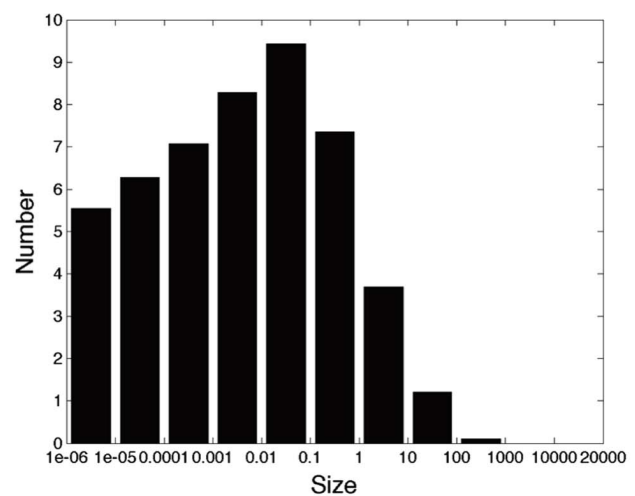

B. Colonies size distribution at the final time

**Figure 3.** *In silico* simulation of the experiment from [56]. A. Time development of the metastatic burden. B. Colonies size distribution at the end time T = 32 days (log-scale on the x-axis).
doi:10.1371/journal.pone.0084249.g003

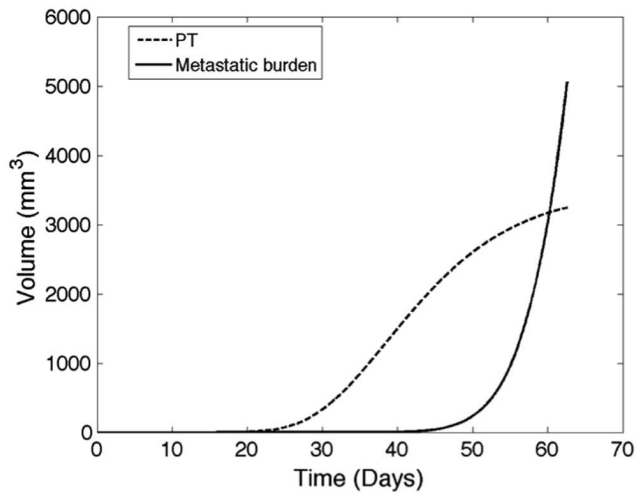
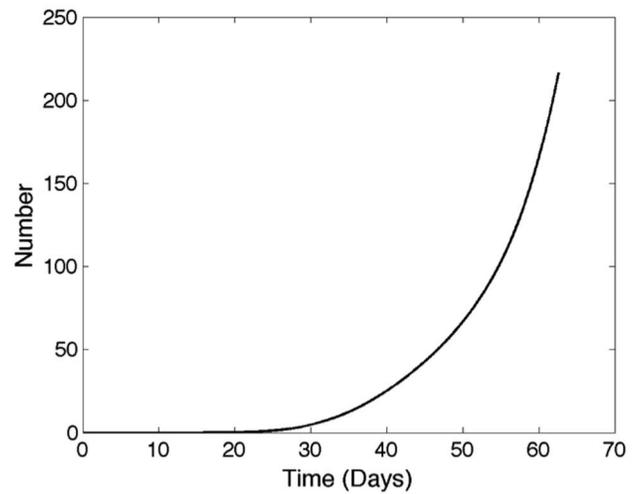

A. PT volume and metastatic burden

B. Total number of metastases

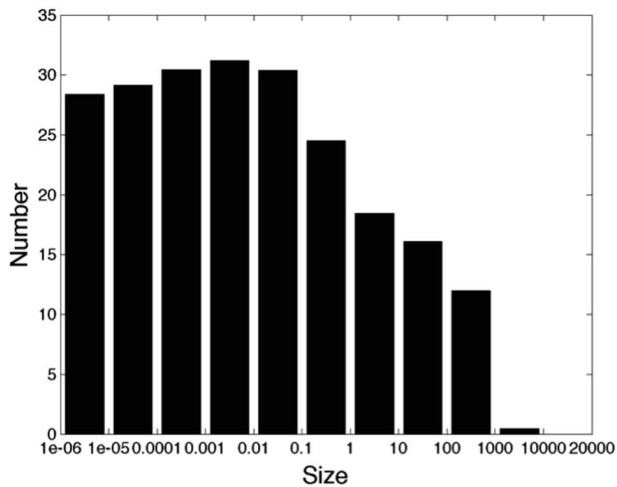
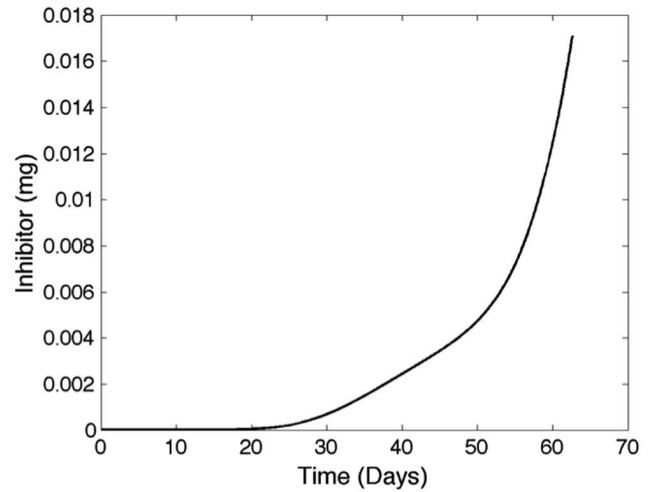

C. End size distribution of metastases

D. Circulating inhibitor

**Figure 4. Simulation of the cancer history from the first cancer cell.** Parameter values are the ones resulting from the fit to the data of [56], reported in Table 1.
doi:10.1371/journal.pone.0084249.g004

In Figure 5, the primary tumor volume, number and total burden of the metastases, the time evolution of the global inhibitor quantity and the size distribution of the metastases at the end time, are plotted.

In this context, the first cancer cell initiates the disease by growing and generating a first pool of metastases, but the metastatic burden then quickly overshadows the growth of the primary lesion (Figure 5.A). The primary tumor reaches a small maximal size of 21.2 mm$^3$ at time 82.9 days (Figure 5.A) and then shrinks due to inhibition of angiogenesis provoked by the distant metastases. There is a slowdown and eventual stabilization of the metastatic burden, with a plateau value of about 2200 mm$^3$. The burden is composed of a large number of metastases (Figure 5.B), most of them being occult micro-metastases as can be seen in the final size distribution (Figure 5.C). This interesting feature of the model simulation could be an *in silico* replicate of the aforementioned situations of cancer without disease [5]. In our model it translates into an asymptotical steady state for the metastatic burden while it is still composed of small lesions. The general dynamics of the metastatic burden results from the balance of two stimulating forces; growth and spread of new individuals, competing with systemic inhibition of angiogenesis. Stimulation depends on the parameters *a* and *b*, which capture the growth process, and *m* and α, which capture spreading. Inhibition depends on *e* and *k*, as well as on *p*, which controls the value of *d*. The present values of the parameters generated long-term stabilization of the mass. The size distribution of the population of secondary tumors at time T = 350 days is revealed to be non trivial, with different numbers in the various size ranges. By 350 days, all the metastases had volume lower than 10 mm$^3$.

In sum, assuming substantial systemic inhibition of angiogenesis, we theoretically obtained an *in silico* replicate of a situation in which an important population of dormant micro-metastases inhibiting each others' growths is present, with a possibly non-lethal final total metastatic burden. This situation was seen to result when a 30-fold higher value for the inhibitor production parameter $p$ was used, compared to the case of growth of a breast cancer line 4T1 extracted from the literature [26], where unlimited expansion of the total metastatic burden was forecast.

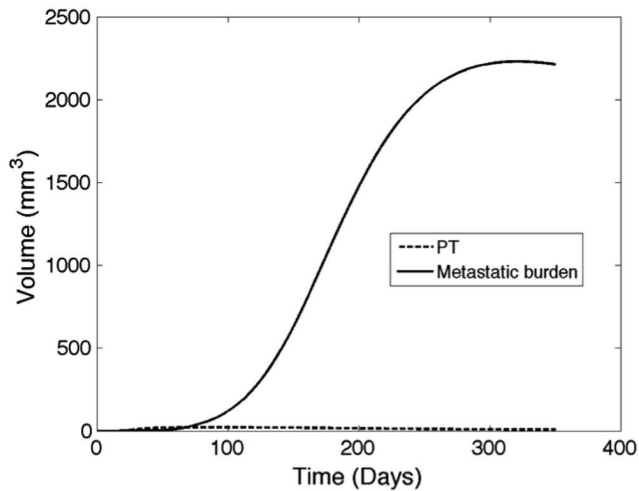
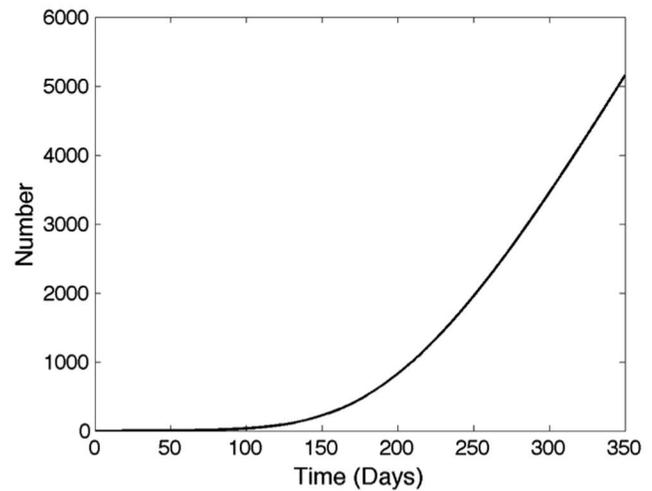
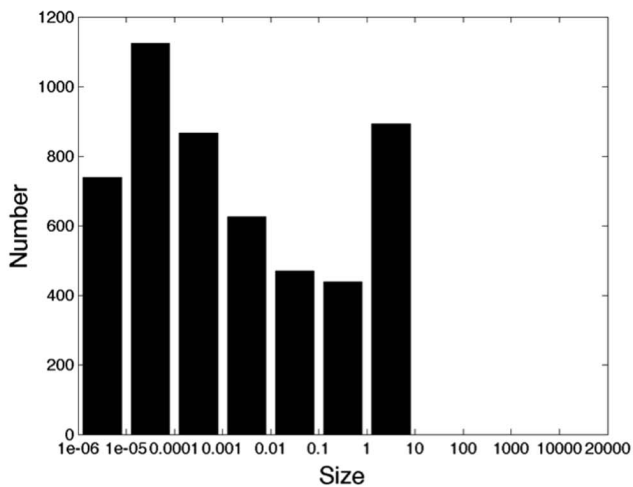
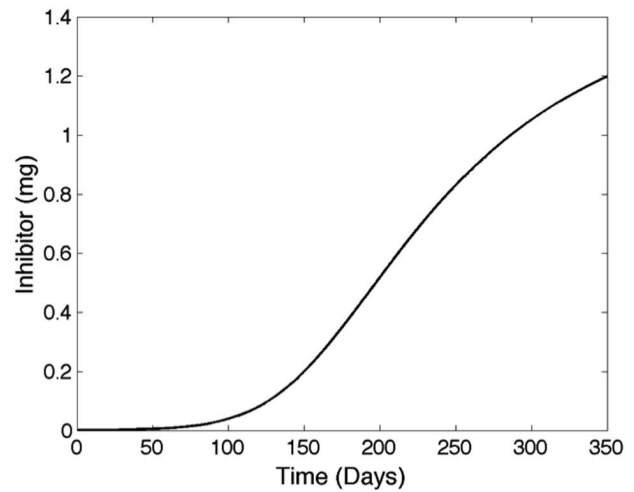

**Figure 5. Large time simulation for large inhibitor production ($p = 2.5 \times 10^{-4}$ mg·mm$^{-3}$·day$^{-1}$, $d = 2.16$ mm$^{-2}$day$^{-1}$).** The model predicts stabilization of the metastatic burden to a situation where the whole metastatic population is in a global dormancy state.
doi:10.1371/journal.pone.0084249.g005

## Discussion

We propose an organism-scale model for the development of a primary tumor and a population of secondary tumors that takes into account systemic inhibiting interactions among tumors due to the release of a circulating angiogenesis inhibitor. The model proves to be able to describe *in vivo* data of primary tumor and metastatic development and allows inference of information not revealed by the experimental data, including the size density distribution of metastases and their total number. Endowed with biologically relevant parameter values, our model is a potentially vital tool for the theoretical study of metastatic dynamics.

It was used here to investigate the whole cancer history from the first cancer cell, and predicted that for the metastatically aggressive 4T1 cell line, metastases would grow unbounded for a physiologically relevant set of parameter values. In this case, the total metastatic burden was found to become larger than the primary tumor mass and probably would be responsible for death of the animal. SIA effects were seen to be negligible in this context. A higher production rate of the inhibitor, by contrast, could theoretically make the primary tumor appearance and growth only a transient event, giving way to a distinct process of tumor development where, due to eventual self-inhibition of angiogenesis at the organism scale, global dormancy is imposed on the entire tumor/metastasis system, stabilizing the cancer disease. Our analysis shows that SIA could conceivably create such a situation, although it would require a very high value of the inhibitor production rate – some 30-fold the value extracted from [26] – which does not appear to be physiological. This suggests that SIA alone is probably not sufficient to induce spontaneous global dormancy and that other processes (such as immune effects) are probably significantly involved. For now, however, our conclusions are limited by the current lack of data on systemic inhibition of tumor development. A study of interactions among multiple tumor implants in controlled immune contexts could shed more light.

Meanwhile, these results inform the human situation by providing elements of explanation for the high prevalence of occult tumors found in autopsy studies. The results as well could inform the consequences of chronic antiangiogenic intervention [58]. As a case in point, progression of cancer disease in individuals having a low production rate of inhibitor might be forestalled, even outside an outright cure, by chronic external administration of supplementary inhibitory agents that could maintain an existing population of tumors in a global dormancy state.

## Acknowledgments

The authors would like to thank Dr. A. d'Onofrio for useful discussions.

## Author Contributions

Analyzed the data: SB. Wrote the paper: SB AG PH. Performed simulations: SB.